\documentclass[conference]{IEEEtran}
\IEEEoverridecommandlockouts
\usepackage{amsmath,amssymb,amsfonts}
\usepackage{algorithmic}
\usepackage{graphicx}
\usepackage{textcomp}
\usepackage{xcolor}
\usepackage{hyperref}
\usepackage{booktabs}
\usepackage{color,soul}
\usepackage{flushend}
\usepackage{multirow}
\usepackage[square, comma, sort&compress,numbers]{natbib}
\def\BibTeX{{\rm B\kern-.05em{\sc i\kern-.025em b}\kern-.08em
    T\kern-.1667em\lower.7ex\hbox{E}\kern-.125emX}}
\begin{document}

\title{Enhancing HDR Video Compression through CNN-based Effective Bit Depth Adaptation}
\author{Chen Feng\textsuperscript{\dag}, Zihao Qi\textsuperscript{\dag}, Duolikun Danier\textsuperscript{\dag}, Fan Zhang\textsuperscript{\dag}, Xiaozhong Xu\textsuperscript{\S}, Shan Liu\textsuperscript{\S} and David Bull\textsuperscript{\dag} \thanks{The authors acknowledge the fundings from China Scholarship Council, the University of Bristol, Tencent (US) and the UKRI
MyWorld Strength in Places Programme.}\\
\textsuperscript{\dag}\textit{Visual Information Laboratory, University of Bristol, Bristol, UK, BS1 5DD}\\
\{chen.feng, zihao.qi, duolikun.danier, fan.zhang, dave.bull\}@bristol.ac.uk\\
\textsuperscript{\S}\textit{Tencent Media Lab, Palo Alto, CA 94306, USA}\\
\{xiaozhongxu, shanl\}@tencent.com}

\maketitle

\begin{abstract}
It is well known that high dynamic range (HDR) video can provide more immersive visual experiences compared to conventional standard dynamic range content. However, HDR content is typically more challenging to encode due to the increased detail associated with the wider dynamic range. In this paper, we improve HDR compression performance using the effective bit depth adaptation approach (EBDA). This method reduces the effective bit depth of the original video content before encoding and reconstructs the full bit depth using a CNN-based up-sampling method at the decoder. In this work, we modify the MFRNet network architecture to enable multiple frame processing, and the new network, multi-frame MFRNet, has been integrated into the EBDA framework using two Versatile Video Coding (VVC) host codecs: VTM 16.2 and the Fraunhofer Versatile Video Encoder (VVenC 1.4.0). The proposed approach was evaluated under the JVET HDR Common Test Conditions using the Random Access configuration. The results show coding gains over both the original VVC VTM 16.2 and VVenC 1.4.0 (w/o EBDA) on JVET HDR tested sequences, with average bitrate savings of 2.9\% (over VTM) and 4.8\% (against VVenC) based on the Bj{\o}ntegaard Delta measurement. The source code of multi-frame MFRNet has been released at \url{https://github.com/fan-aaron-zhang/MF-MFRNet}.
\end{abstract}


\section{Introduction}
\label{sec:intro}

High dynamic range (HDR) formats can offer improved audience experiences with extended perceived depth through increased bit depth, color gamut, and screen bright-to-dark range \cite{bull2021,reinhard2010high}. HDR content typically exhibits increased spatial detail but, as a consequence, is more difficult to encode. The standard parameters for HDR content production and transmission are defined in ITU-R BT.2100 \cite{sector2018image}, based on two methods, Perceptual Quantisation (PQ) and Hybrid Log-Gamma (HLG). Both of these are supported by recent MPEG video coding standards: High Efficiency Video Coding (HEVC) \cite{hevc} and Versatile Video Coding (VVC) \cite{vvc}.

The latest VVC test model (VTM) includes several coding tools specifically designed for HDR content; these include luma mapping with chroma scaling and luma-adaptive deblocking \cite{JVETVVCencoder16}. In the wider research community, many other approaches have also been proposed to enhance HDR video compression performance, typically focusing on perceptual quantisation \cite{zhang2015high,liu2017adaptive}, rate-distortion optimisation \cite{mir2017efficient} and solving chroma leakage problems \cite{franccois2019high}. 


In recent years, deep learning techniques have been increasingly exploited to optimise video coding performance, mainly focused on standard dynamic range (SDR) content. Notable examples include CNN-based intra-frame prediction \cite{cui2018convolutional,kuanar2018fast}, inter-frame prediction \cite{zhang2019advanced,wang2018neural}, post-processing \cite{ma2020mfrnet,PPMultiMedia,ma2020cvegan} and loop filters \cite{wang2018dense,li2019deep}, and resolution adaptation \cite{afonso2019video,ma2019perceptually}. Similar methods have also been extended to HDR video compression based on CNN-inspired super-resolution \cite{umeda2018hdr}, pre-processing \cite{ki2020learning} and post-processing \cite{eilertsen2017hdr}.

Among these deep learning-based coding approaches, the effective bit depth adaptation (EBDA) coding framework has been reported to offer consistent coding gains on SDR content when integrated with HEVC and VVC \cite{zhang2019enhanced,zhang2019vistra2,ma2020gan}. In this method, the CNN-based effective bit depth (EBD) up-sampling operation is similar to the inverse tone mapping process for HDR content \cite{banterle2006inverse,lee2018deep,rana2019deep}, which converts low bit depth content to higher precision formats. However, the EBDA method has not previously been evaluated in the context of HDR video compression.

\begin{figure*}[ht]
\centering
 \includegraphics[width=\linewidth]{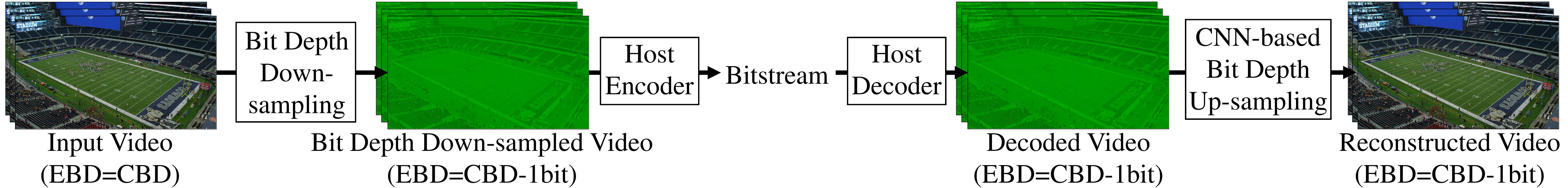}
\caption{Diagram of the EBDA coding framework.\label{fig:framework_single_Chen}}
\end{figure*}

\begin{figure}[ht]
\centering
\includegraphics[width=\linewidth]{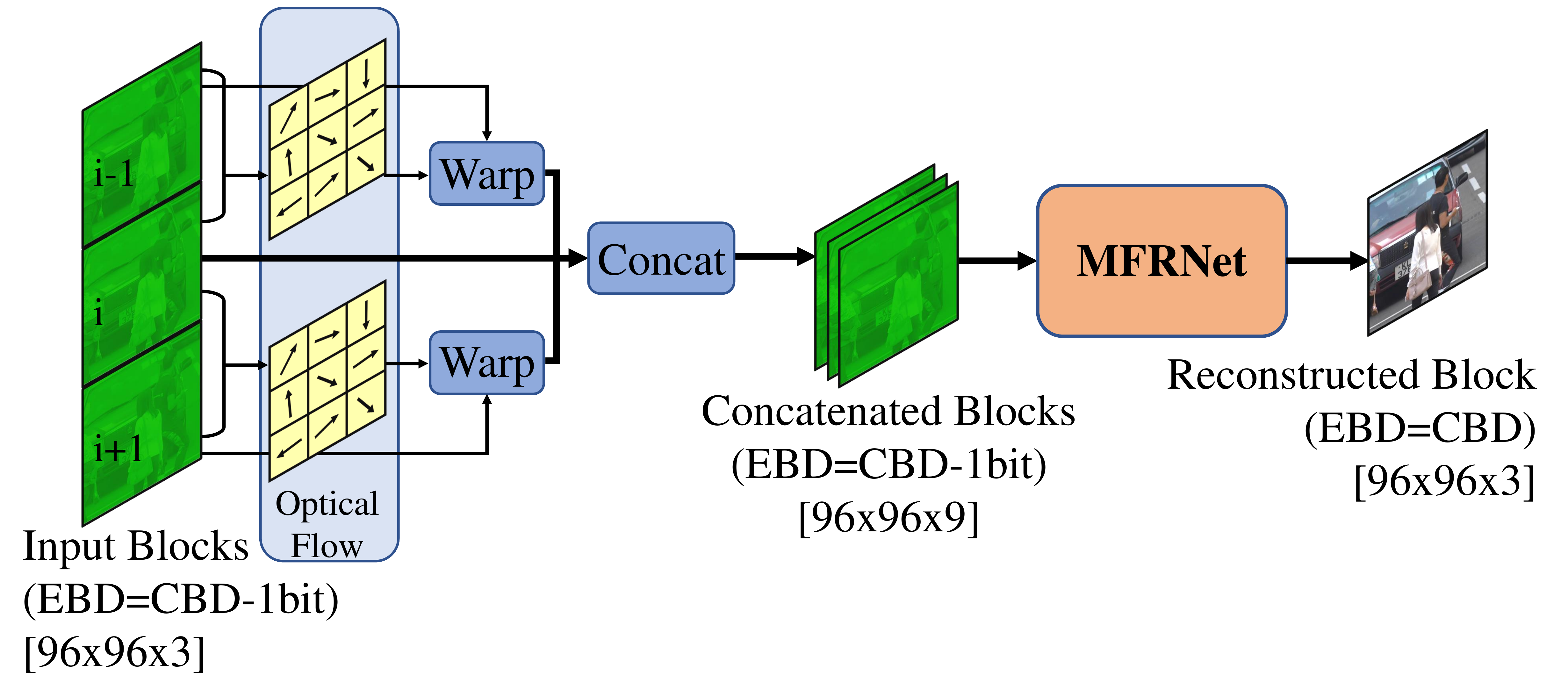}
\caption{Network architecture of the employed multi-frame MFRNet.\label{fig:architecture}}
\end{figure}

In this paper, we extend the application of the EBDA coding approach \cite{zhang2019vistra2} from SDR content to HDR, and employ a modified CNN architecture, multi-frame MFRNet \cite{ma2020mfrnet}, for EBD up-sampling. The new multi-frame MFRNet uses a similar backbone to that in MFRNet, which contains Multi-level Feature review Residual dense Blocks (MFRB) in a cascaded structure. This has been previously reported to provide superior performance for post-processing of compressed video content. Inspired by the success of pre-warping for texture synthesis~\cite{zhang2011parametric}, we changed the original MFRNet architecture to enable multiple (three) frames as input and perform spatio-temporal EBDA based on a low complexity optical flow based warping approach. The new EBDA algorithm has been integrated within both the VVC VTM 16.2 reference software and VVenC 1.4.0 (VVC open-source encoder implementation) and tested using the Joint Video Exploration Team (JVET) HDR Common Test Conditions (CTC) in the Random Access (RA) configuration \cite{bossen2018jvet}. The results show that the proposed method outperforms the original VVC VTM 16.2 and VVenC 1.4.0 on JVET HDR test sequences, with average coding gains of 2.9\% and 4.8\% respectively.



The remainder of this paper is organised as follows. Section \ref{sec:algorithm} describes the employed EBDA coding framework,  the new CNN architecture employed for EBD up-sampling and the network training process. The compression results and complexity analysis are then reported in Section \ref{sec:results}. Finally, Section \ref{sec:conclusion} concludes the paper and proposes future work directions.

\section{Proposed Algorithm}
\label{sec:algorithm}

The employed EBDA coding framework is illustrated in Fig. \ref{fig:framework_single_Chen}. This reduces the EBD before encoding, and restores the full EBD video frames at the decoder using a CNN-based EBD up-sampling approach. EBD defines the actual bit depth that is utilised to represent the video content; this is different from CBD (coding bit depth) i.e. the pixel bit depth used for video coding. In the VVC VTM software, CBD is also denoted as \texttt{InternalBitDepth}. 

In the framework shown in Fig. \ref{fig:framework_single_Chen}, the CBD remains the same throughout the whole coding workflow. The bit depth reduction is 1 bit, the same as adopted in previous work~\cite{zhang2019vistra2}. The EBD down-sampling operation is implemented using simple bit-shifting. Moreover, to allow a meaningful comparison with the anchor codecs (VVC VTM 16.2 and VVenC 1.4.0 w/o EBDA in this case) within a similar bit rate range, a quantisation parameter (QP) offset of -6 is applied during encoding. This value is based on the empirical observation described in \cite{afonso2017low,zhang2019enhanced}.

\subsection{The modified CNN architecture}

In this work, a modified CNN architecture, multi-frame MFRNet \cite{ma2020mfrnet}, has been employed for EBD up-sampling (illustrated in Fig. \ref{fig:architecture}). The input to this network consists of three $3\times 96 \times 96$ YCbCr 4:4:4 image blocks from three consecutive reconstructed frames (at the same spatial location) with reduced EBD (CBD-1bit), while the output is a $3\times96\times96$ image block with the same format but targeting the uncompressed full EBD (the same as CBD) counterpart in the intermediate frame. The original MFRNet has been reported to provide better enhancement performance when used for post-processing and in-loop filtering \cite{ma2020mfrnet}, compared to other simpler network architectures such as MSRResNet \cite{zhang2019vistra2,zhang2019enhanced}. We adapted the original MFRNet to accept a multi-frame (three in this case) patch as network input, and the motion information between neighbouring frames is calculated using a low complexity optical flow method \cite{kroeger2016fast}, and fed into the network for frame warping (the intermediate frame is used as the warping benchmark). By performing such frame alignment, the model is able to access additional information about each pixel in multiple consecutive frames. This also avoids introducing 3D convolutional layers into the network to perform spatio-temporal filtering, which typically presents very high computational complexity. The ablation study in Section \ref{sec:results} validates this design, demonstrating the contribution of multi-frame processing to the overall coding gain.
\begin{table*}[t]
\centering
\caption{Compression results of the proposed HDR EBDA approach with \textbf{VTM 16.2} and \textbf{VVenC 1.4.0} on the JVET HDR CTC tested sequences.}
\small
\resizebox{\linewidth}{!}{
\begin{tabular}{l|c|c|c|c|c|c|c|c}
\toprule
 \textbf{}&  \multicolumn{4}{c|}{\textbf{VTM 16.20}}&  \multicolumn{4}{c}{\textbf{VVenC 1.4.0}}\\\midrule 
 \multirow{2}{*}{\textbf{Sequence}}& \multicolumn{2}{c|}{\textbf{MFRNet}}& \multicolumn{2}{c|}{\textbf{Multi-frame MFRNet}}& \multicolumn{2}{c|}{\textbf{MFRNet}}& \multicolumn{2}{c}{\textbf{Multi-frame MFRNet}}\\
 \cmidrule{2-9}
& \textbf{BD-Rate}& \textbf{BD-PSNR}& \textbf{BD-Rate}& \textbf{BD-PSNR}& \textbf{BD-Rate}& \textbf{BD-PSNR}& \textbf{BD-Rate}& \textbf{BD-PSNR}\\
\midrule 
H1-BalloonFestival&-1.1\% & +0.06dB & -1.7\% & +0.08dB &-10.5\% & +0.50dB &-10.7\% & +0.52dB \\
H1-Cosmos1&-2.7\% & +0.08dB &-2.5\% & +0.08dB &-1.4\%& +0.05dB &-1.8\% & +0.06dB\\
H1-Hurdles&-6.9\% & +0.25dB &-7.4\% & +0.27dB &-2.6\% &+0.07dB  &-2.6\% & +0.08dB\\
H1-Market3&-4.8\% & +0.19dB &-5.0\% & +0.20dB &-5.1\% & +0.20dB&-5.0\% & +0.20dB\\
H1-Starting&-1.4\% & +0.10dB &-2.4\% & +0.13dB &-7.9\% & +0.34dB &-8.2\% & +0.37dB\\
H1-SunRise&+2.9\% & -0.07dB &+1.7\% & -0.04dB &-9.3\% & +0.28dB &-10.1\% & +0.31dB\\
\midrule \textbf{Class H1 } & \textbf{-2.3\% }& \textbf{+0.10dB}
& \textbf{-2.9\%}&\textbf{+0.12dB}& \textbf{-6.1\%}
& \textbf{+0.24dB}&\textbf{-6.4\%}&\textbf{+0.26dB }\\
\midrule
H2-DayStreet&-1.7\% & +0.07dB &-2.0\% & +0.08dB &-0.1\% & +0.03dB &-0.3\% & +0.03dB\\
H2-PeopleInShopping&-3.0\% & +0.15dB &-3.4\% &+0.16dB &-0.1\% & +0.05dB &-0.4\% & +0.05dB\\
H2-SunsetBeachs&-2.7\% & +0.10dB &-3.1\% & +0.11dB &-3.4\% & +0.12dB &-3.8\% & +0.14dB\\
\midrule\textbf{Class H2 } & \textbf{-2.5\% }& \textbf{+0.10dB}
&\textbf{ -2.8\% }& \textbf{+0.12dB}&\textbf{-1.2\% }& \textbf{+0.7dB}
& \textbf{-1.5\%}&\textbf{+0.7dB} \\
\midrule   \textbf{Overall} & \textbf{-2.4\% }&
\textbf{+0.10dB} & \textbf{-2.9}\% & \textbf{+0.12dB} & \textbf{-4.5\% }& \textbf{+0.18dB}
& \textbf{-4.8\%}&\textbf{+0.20dB} \\
\bottomrule
\end{tabular}
}
\label{tab1}
\end{table*}

\subsection{Network training}

In this work, we used a large video database, BVI-DVC \cite{ma2020bvi} to train the multi-frame MFRNet for EBD up-sampling. This database has been used by MPEG JVET to train neural network-based coding tools for VVC \cite{jvett2006}.  It contains 800 source sequences with diverse and representative content, all of which have 64 frames with 10 bit YCbCr 4:2:0 format at four different spatial resolutions (270p, 540p, 1080p and 2160p). The EBD of these 800 videos is down-sampled from 10 to 9 bits using bit-shifting and they are then compressed using VVC VTM 16.2 and VVenC 1.4.0 based on the JVET-CTC Random Access (RA) configuration with four initial base QP values, 22, 27, 32 and 37. A fixed QP offset of -6 is applied in the encoding process as mentioned above. The compressed video frames and their original counterparts were randomly selected, segmented into three 3$\times$96$\times$96 image blocks (from three consecutive frames) and converted into YCbCr 4:4:4 format. After data augmentation (through block rotation), there are approximately 178,000 pairs of image blocks for each QP group.

The training of the multi-frame MFRNet is based on TensorFlow (version 1.8.0) and employs the $\ell1$ loss function. Other training parameters include the ADAM optimiser \cite{kingma2014adam} with hyper-parameters of $\beta_1$=0.9 and $\beta_2$=0.999, 0.0001 learning rate, 0.1 learning rate decay per 100 epochs, 200 total training epochs, and batch size of 16. Four CNN models were trained for different QP groups (22, 27, 32 and 37) and each host codec (VTM or VVenC) and then used in the evaluation stage according to the used base QP values (before applying the offset):

\begin{equation}
\label{eq4}
{M_\mathrm{CNN}}=\left\{
\begin{aligned}
{M_1} & , & {\rm QP_\mathrm{base}}\leq 24.5\\
{M_2} & , & {\rm 24.5 < QP_\mathrm{base}} \leq 29.5\\
{M_3} & , & {\rm 29.5 < QP_\mathrm{base}} \leq 34.5\\
{M_4} & , & {\rm QP_\mathrm{base}} > 34.5\\
\end{aligned}
\right.
\end{equation}
 
\section{Results and Discussion}
\label{sec:results}

The proposed EBDA framework for HDR has been integrated into VVC VTM 16.2 and VVenC 1.4.0 as host codecs and evaluated under the JVET HDR Common Test Conditions \cite{bossen2018jvet}, on nine HDR test sequences at spatial resolutions of 3840$\times$2160, 1920$\times$1080 and 1920$\times$856. All original sequences are in 10 bit YCbCr 4:2:0 format obtained based on PQ (Class H1) or HLG (Class H2) transformation functions. None of these sequences have been included in the CNN training. The proposed approach was compared to the original VVC VTM 16.2 and VVenC 1.4.0 using the Bj{\o}ntegaard Delta \cite{BD} measurement (BD-rate) based on PSNR (luma channel only). The encoding and decoding operations were executed using a shared cluster (BlueCrystal Phase 4) computer \cite{BC4} at the University of Bristol. Each node includes an NVIDIA P100 GPU, two 14 core 2.4 GHz Intel E5-2680 v4 (Broadwell) CPUs, and 128 GB of RAM.

\begin{figure*}[ht]
\centering
\scriptsize
\centering
\begin{minipage}[b]{0.245\linewidth}
\centering
\centerline{\includegraphics[height=0.8\linewidth]{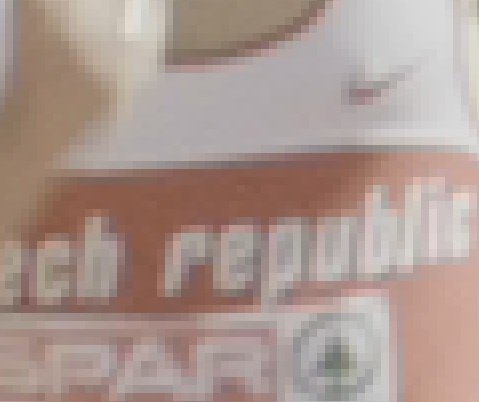}}
(a) Original \\ \ \ 
\end{minipage}
\begin{minipage}[b]{0.245\linewidth}
\centering
\centerline{\includegraphics[height=0.8\linewidth]{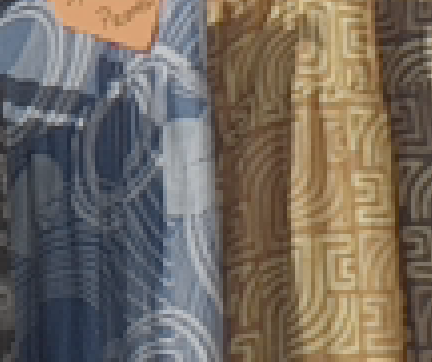}}
(b) Original \\ \ \ 
\end{minipage}
\begin{minipage}[b]{0.245\linewidth}
\centering
\centerline{\includegraphics[height=0.8\linewidth]{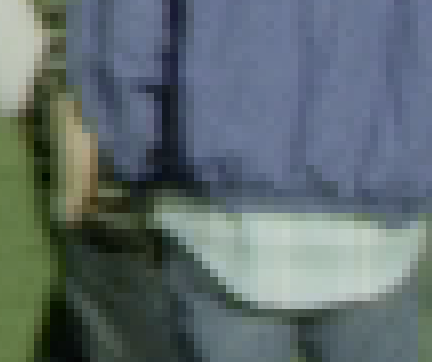}}
(c) Original \\ \ \ 
\end{minipage}
\begin{minipage}[b]{0.245\linewidth}
\centering
\centerline{\includegraphics[height=0.8\linewidth]{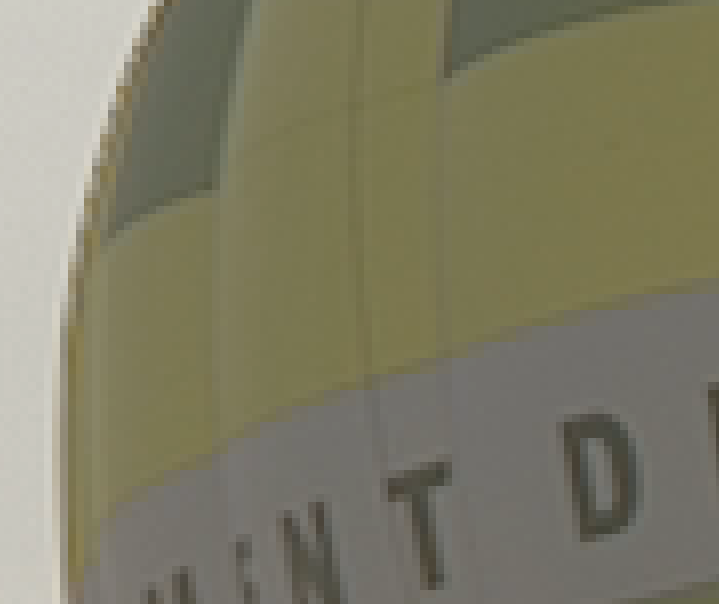}}
(d) Original \\ \ \ 
\end{minipage}

\begin{minipage}[b]{0.245\linewidth}
\centering
\centerline{\includegraphics[height=0.8\linewidth]{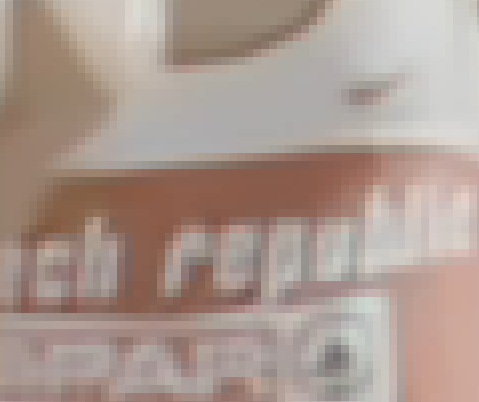}}
(e) VTM 16.2, QP=32 \\  \ \ 
\end{minipage}
\begin{minipage}[b]{0.245\linewidth}
\centering
\centerline{\includegraphics[height=0.8\linewidth]{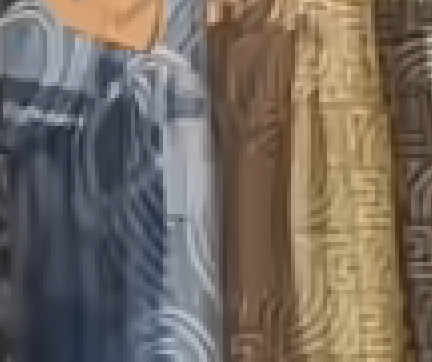}}
(f) VTM 16.2, QP=37 \\  \ \ 
\end{minipage}
\begin{minipage}[b]{0.245\linewidth}
\centering
\centerline{\includegraphics[height=0.8\linewidth]{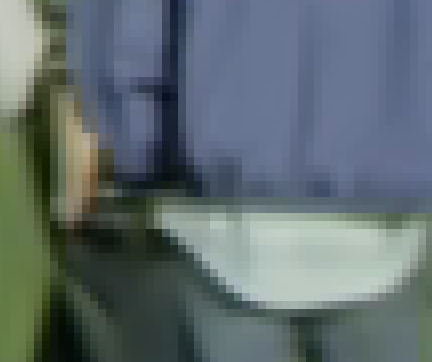}}
(g) VVenC 1.4.0, QP=32 \\ \ \ 
\end{minipage}
\begin{minipage}[b]{0.245\linewidth}
\centering
\centerline{\includegraphics[height=0.8\linewidth]{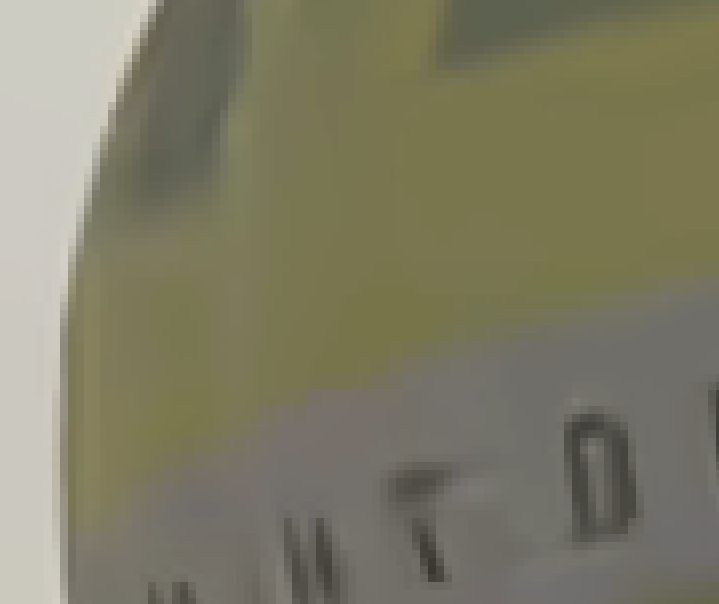}}
(h) VVenC 1.4.0, QP=37 \\ \ \ 
\end{minipage}


\begin{minipage}[b]{0.245\linewidth}
\centering
\centerline{\includegraphics[height=0.8\linewidth]{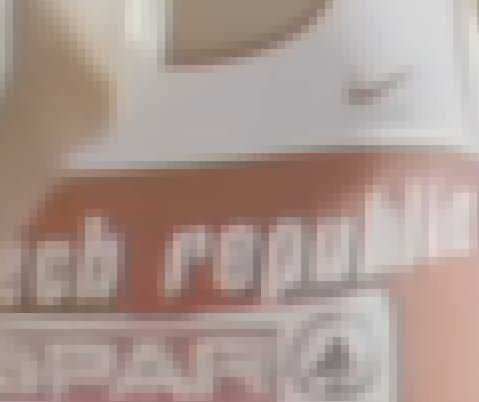}}
(i) Multi-frame MFRNet, QP=26 \\  \ \ 
\end{minipage}
\begin{minipage}[b]{0.245\linewidth}
\centering
\centerline{\includegraphics[height=0.8\linewidth]{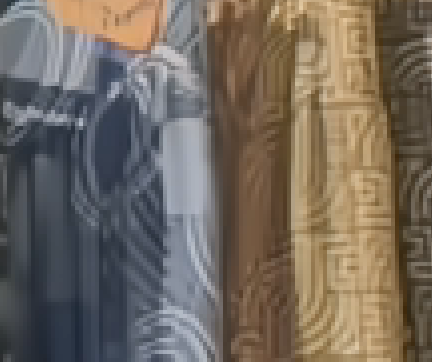}}
(j) Multi-frame MFRNet, QP=31 \\  \ \ 
\end{minipage}
\begin{minipage}[b]{0.245\linewidth}
\centering
\centerline{\includegraphics[height=0.8\linewidth]{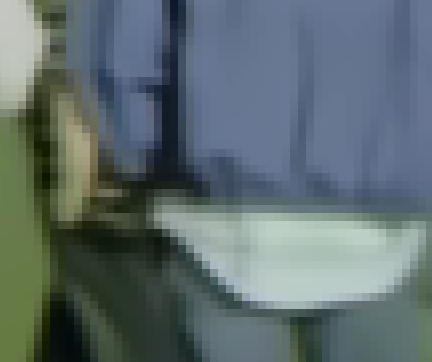}}
(k) Multi-frame MFRNet, QP=26 \\ \ \ 
\end{minipage}
\begin{minipage}[b]{0.245\linewidth}
\centering
\centerline{\includegraphics[height=0.8\linewidth]{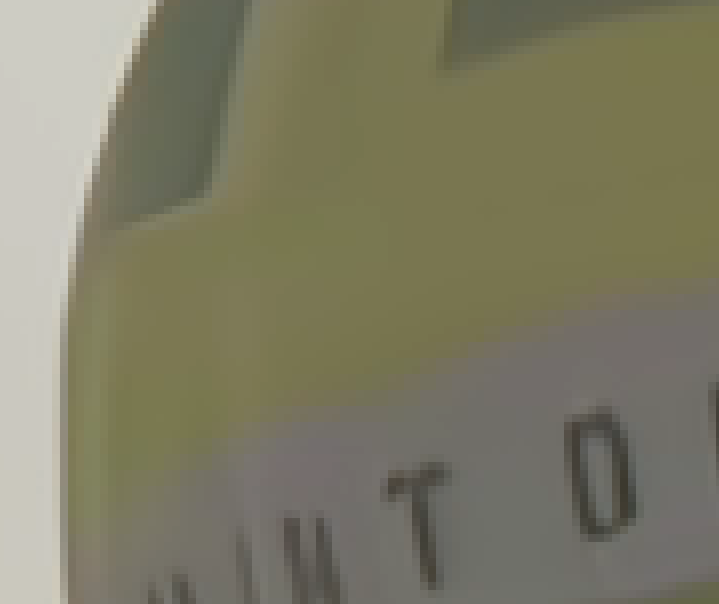}}
(l) Multi-frame MFRNet, QP=31 \\ \ \ 
\end{minipage}

\caption{Example blocks of the reconstructed frames generated by the anchor VTM 16.2, VVenC 1.4.0, and EBDA (each pair has bitstreams with similar bit rates). These blocks are from the 1st frames of Hurdles (Column 1), Market3 (Column 2), BalloonFestival (Column 3) and H1SunRise (Column 4) sequences.\label{fig:perceptual}}
\end{figure*}

TABLE \ref{tab1} summarises the compression performance of the proposed CNN-based EBDA framework benchmarked on the original VVC VTM 16.2 and VVenC 1.4.0. It is noted that evident coding gains have been achieved by the proposed method (Multi-frame MFRNet) for most HDR test sequences in both classes, with the average BD-rate of -2.9\% and -4.8\% against VTM and VVenC respectively based on PSNR. The only exception is on the SunRise sequence in the Class H1, where a small coding loss is obtained against VTM (+1.7\% on PSNR).

To evaluate the effectiveness of the proposed multi-frame MFRNet architecture, we have also generated compression results (also shown in TABLE \ref{tab1}) for comparison using the original MFRNet (optimised on the same training database). It can be observed that the average coding gains (in terms of BD-rate) when using multi-frame MFRNet is higher than that for the original (single frame) MFRNet, and the improvement is consistent among various test sequences.

Alongside the quantitative results, we also included a visual comparison (shown in Fig. \ref{fig:perceptual}) between the original VVC compressed content and the final reconstructed frames by EBDA (using Multi-frame MFRNet) when the bitrates are similar. It can be observed that example blocks generated by the proposed approach exhibit improved perceptual quality compared to the anchors (VVC VTM 16.2 and VVenC 1.4.0) with fewer compression artefacts and increased textural detail.  

Finally, we calculated the encoder and decoder complexity figures of the proposed method, benchmarked on the original VVC VTM and VVenC. The average encoding time is 0.98 and 1.02 times that of the original VVC VTM 16.2 encode and VVenC 1.4.0, respectively, while the relative complexity of the EBDA decoder is 191.14 and 172.21 times that of VTM and VVenC without memory optimisation or parallel processing. These figures are based on all the test sequences and QP values. The high decoding complexity is primarily due to the employed multi-frame CNN-based bit-depth up-sampling operation. Our future work should focus on significantly reducing the computational complexity of the proposed multi-frame MFRNet.

\section{Conclusion}
\label{sec:conclusion}

This paper presents an HDR video compression enhancement method based on effective bit depth adaptation. This employs a modified CNN architecture, multi-frame MFRNet, for EBD up-sampling. The proposed method has been integrated into both the VVC VTM 16.2 and VVenC 1.4.0 reference codecs, and evaluated under the JVET HDR CTC using the Random Access configuration. The proposed method outperforms the original VTM 16.2 and VVenC 1.4.0 on most tested HDR sequences, with average bitrate savings of 2.9\% and 4.8\% respectively. Future work will focus on the complexity reduction of CNN computation.

\small
\bibliographystyle{IEEEtran}
\bibliography{IEEEexample}
\end{document}